\documentclass{birkjour}
\usepackage{tikz}
 \theoremstyle{definition}
 
 \theoremstyle{remark}

 \numberwithin{equation}{section}

\begin{document}

%
%
%
%
%
%
%
%
%

\title[Nonassociative generalization of supersymmetry]
 {Nonassociative generalization of supersymmetry}

\author[Vladimir Dzhunushaliev]{Vladimir Dzhunushaliev}

\address{%
Dept. Theor. and Nucl. Phys., KazNU, Almaty, 050040, Kazakhstan; \\
IETP, Al-Farabi Kazakh National University, Almaty, 050040, Kazakhstan }

\email{v.dzhunushaliev@gmail.com}




\date{January 1, 2004}
\dedicatory{To my wife Nina and childrens: Danil and Natalia}

\begin{abstract}
A nonassociative generalization of supersymmetry is studied, where supersymmetry generators are considered to be the nonassociative ones. Associators for the product of three and four multipliers are defined. Using a special choice of the parameters, it is shown that  the associator of the product of four supersymmetry generators is connected with the angular momentum operator. The connection of operator decomposition to the hidden variables theory and alternative quantum mechanics is discussed.
\end{abstract}

\maketitle
\section{Introduction}
Supersymmetry is one of the important parts of the paradigm of modern physics. Nonassociativity is a rare guest in modern theoretical physics. Nevertheless, Ref. \cite{okubo1995} offers many examples in the use of
nonassociativity in physics. At the present time, the  following nonassociative topics are under consideration: the classification of quaternionic and octonionic realizations of Clifford algebras and spinors \cite{Carrion:2003ve}; nonassociative octonionic ternary gauge field theories based on a ternary bracket  \cite{Castro:2012zzb}; octonionic electrodynamics and the Dirac equation \cite{Gogberashvili:2005xb,Gogberashvili:2005cp}; gauge theory on nonassociative spaces \cite{Beggs:2005zt}; nonassociative geometry and a discrete spacetime \cite{Sabinin:2001pd}; the Standard model within nonassociative geometry~\cite{Wulkenhaar:1996at}.

Here we would like to show that the supersymmetry has a nonassociative generalization; to give exact definitions of three and four associators; to show that for a special definition of a four associator, this associator will be connected with the angular momentum operator; and to discuss some interesting problems arising from the obtained decomposition of quantum operators. In this connection, it is necessary to note a four-dimensional nontrivial extension of the Poincar\'e algebra, distinct from the
supersymmetry -- the so-called fractional supersymmetry \cite{fsusy,Mohammedi:2003qx}. In supersymmetry, the extensions of the Poincar\'e algebra are obtained from a square root of the translations, $QQ \sim P$. In cubic supersymmetry, new algebras occur as a result of using of associator $\left[ Q, Q, Q \right] \sim P$, see Ref.~\cite{Rausch de Traubenberg:2003bi}.

Another interesting approach to extend supersymmetry is a ternary algebra. Ternary algebras may give a unified description of Lie algebras and superalgebras \cite{Bars:1978yx}. In Ref. \cite{Castro:2014tda} the nonassociative and noncommutative octonionic ternary gauge ﬁeld theory
 based on a ternary-bracket structure involving the octonion algebra is considered. The ternary bracket obeys the Nambu fundamental identity
and was developed by Yamazaki \cite{Yamazaki:2008gg}.

\section{The simplest supersymmetry algebra}

The simplest supersymmetry algebra is defined as
\begin{eqnarray}
  \left\{
     Q_a , Q_{\dot a}
  \right\} &=& Q_a Q_{\dot a} + Q_{\dot a} Q_a =
  2 \sigma^\mu_{a \dot a} P_\mu,
\label{2-10}  \\
  \left\{
     Q_a , Q_b
  \right\} &=& \left\{
     Q_{\dot a} , Q_{\dot b}
  \right\} = 0,
\label{2-20} \\
  \left[ Q_a , P_\mu \right] &=& \left[ Q_{\dot a} , P_\mu \right] = 0,
\label{2-30}\\
  \left[ P_\mu , P_\nu \right] &=& 0,
\label{2-40}
\end{eqnarray}
where $Q_a, Q_{\dot a}$ are supersymmetry generators;
$P_\mu = -i \partial_\mu$; $\mu = 0,1,2,3$; $a=1,2$;
$\dot a = \dot 1, \dot 2$, and the Pauli matrices
$\sigma^\mu_{a \dot a}, \sigma_\mu^{a \dot a}$ are
\begin{eqnarray}
  \sigma^\mu_{a \dot a} &=& \left\{
    \left(
      \begin{array}{cc}
        1 & 0 \\
        0 & 1 \\
      \end{array}
    \right),
    \left(
    \begin{array}{cc}
        0 & 1 \\
        1 & 0 \\
      \end{array}
    \right),
    \left(
    \begin{array}{cc}
        0 & -\imath \\
        \imath & 0 \\
      \end{array}
    \right),
    \left(
    \begin{array}{cc}
        1 & 0 \\
        0 & -1 \\
      \end{array}
    \right)
  \right\}
\label{2-50}\\
  \sigma_\mu^{a \dot a} &=& \left\{
    \left(
      \begin{array}{cc}
        1 & 0 \\
        0 & 1 \\
      \end{array}
    \right),
    \left(
    \begin{array}{cc}
        0 & 1 \\
        1 & 0 \\
      \end{array}
    \right),
    \left(
    \begin{array}{cc}
        0 & \imath \\
        -\imath & 0 \\
      \end{array}
    \right),
    \left(
    \begin{array}{cc}
        1 & 0 \\
        0 & -1 \\
      \end{array}
    \right)
  \right\},
\label{2-60}
\end{eqnarray}
with the following relations,
\begin{equation}\label{2-70}
  \sigma_\mu^{a \dot a} \sigma^\nu_{a \dot a} = 2 \delta_\mu^\nu, \quad
  \sigma_\mu^{a \dot a} \sigma^\mu_{b \dot b} =
  2 \delta_b^a \delta_{\dot b}^{\dot a}, \quad \imath^2 = -1.
\end{equation}
An inverse relation for \eqref{2-10} is
\begin{equation}\label{2-80}
  P_\mu = \frac{1}{4} \sigma_\mu^{a \dot a} \left\{
    Q_a, Q_{\dot a}
  \right\}.
\end{equation}
The main goal of this Letter is to show that one can generalize supersymmetry in such a way that supergenerators $Q_a, Q_{\dot a}$ become nonassociative ones.

\section{Nonassociative generalization of the simplest supersymmetry algebra}

Let us define an associator as follows:
\begin{equation}\label{3-10}
  \left[ x,y,z \right] = \left( x y \right) z - x \left( y z \right).
\end{equation}
Let us now define associators for supersymmetry generators. First, we define an associator for the product of three generators,
\begin{equation}
  \left[ Q_x, Q_y, Q_z \right] = 0,
\label{3-20}
\end{equation}
where the triple $x,y,z$ is any combination of dotted and undotted indices
$a, \dot a$. We define an associator for the product of four generators  as
\begin{eqnarray}
  \left[ Q_a, Q_b, \left( Q_x Q_y \right) \right] &=&
  \left[ Q_{\dot a}, Q_{\dot b}, \left( Q_x Q_y \right) \right] = 0,
\label{3-30}\\
  \left[ Q_a, Q_{\dot a}, \left( Q_b Q_{\dot b} \right) \right] &=&
  \alpha_{a \dot a} \left\{ Q_b, Q_{\dot b} \right\} +
  \beta_{b \dot b} \left\{ Q_a, Q_{\dot a} \right\},
\label{3-40}\\
  \left[ Q_{\dot a}, Q_a, \left( Q_b Q_{\dot b} \right) \right] &=&
  \gamma_{a \dot a} \left\{ Q_b, Q_{\dot b} \right\} +
  \delta_{b \dot b} \left\{ Q_a, Q_{\dot a} \right\},
\label{3-45}\\
  \left[ \left( Q_a Q_{b} \right), Q_x, Q_y \right] &=&
  \left[ \left( Q_{\dot a} Q_{\dot b} \right), Q_x, Q_y \right] = 0,
\label{3-42}\\
  \left[ \left( Q_a Q_{\dot a} \right), Q_b, Q_{\dot b} \right] &=&
  \tilde \alpha_{a \dot a} \left\{ Q_b, Q_{\dot b} \right\} +
  \tilde \beta_{b \dot b} \left\{ Q_a, Q_{\dot a} \right\},
\label{3-44}\\
  \left[ \left( Q_a Q_{\dot a} \right), Q_{\dot b}, Q_b \right] &=&
  \tilde \gamma_{a \dot a} \left\{ Q_b, Q_{\dot b} \right\} +
  \tilde \delta_{b \dot b} \left\{ Q_a, Q_{\dot a} \right\},
\label{3-44a}\\
  \left[ Q_a, \left( Q_b Q_x \right), Q_y \right] &=&
  \left[ Q_{\dot a}, \left( Q_{\dot b} Q_x \right), Q_y \right] = 0,
\label{3-46}\\
  \left[ Q_a, \left( Q_{\dot a} Q_{\dot b} \right), Q_{b} \right] &=&
  \left[ Q_{\dot a}, \left( Q_{a} Q_{b} \right), Q_{\dot b} \right] = 0,
\label{3-49}\\
  \left[ Q_a, \left( Q_{\dot a} Q_b \right), Q_{\dot b} \right] &=&
  \tilde {\tilde \alpha}_{a \dot a} \left\{ Q_b, Q_{\dot b} \right\} +
  \tilde {\tilde \beta}_{b \dot b} \left\{ Q_a, Q_{\dot a} \right\},
\label{3-48}
\end{eqnarray}
where the pair $x,y$ is any combination of dotted and undotted indices
$a, \dot a$; $\alpha_{a \dot a}$ and $\beta_{b \dot b}$ are complex numbers.

Using pentagon identity
\begin{equation}
\begin{tikzpicture}[line width=2pt, scale=0.6]
\draw[->]
(-12,-8.5) node [below,outer sep=0,inner sep=1,minimum size=10] (v1)
{
$\left(Q_x \left( Q_yQ_z \right) \right) Q_u$
} --
(-7,-8.5) node [below,outer sep=0,inner sep=1,minimum size=10]
{
$Q_x \left( \left( Q_y Q_z \right) Q_u \right)$
} ;
\draw[->]
(-7,-8.5) node [below,outer sep=0,inner sep=1,minimum size=10]
{ } --
(-6,-5) node [right,outer sep=0,inner sep=1,minimum size=10]
{ } ;
\draw[<-]
(-6,-5) node [right,outer sep=0,inner sep=1,minimum size=10]
{
$Q_x \left( Q_y \left( Q_z Q_u \right) \right)$
} --
(-9.5,-2.5)node [above,outer sep=0,inner sep=1,minimum size=10]
{ } ;
\draw[<-]
(-9.5,-2.5)node [above,outer sep=0,inner sep=1,minimum size=10]
{
$\left(Q_x Q_y\right) \left( Q_z Q_u \right)$
} --
(-13,-5) node [left,outer sep=0,inner sep=1,minimum size=10]
{ } ;
\draw[->]
(-13,-5) node [left,outer sep=0,inner sep=1,minimum size=10]
{
$\left( \left(Q_x Q_y\right) Q_z \right) Q_u$
} --
(-12,-8.5) node [below] (v1)
{ };
\end{tikzpicture}
\label{3-49b}
\end{equation}
one can bind
$\alpha_{a \dot a}, \tilde{\alpha}_{a \dot a}, \tilde{\tilde \alpha}_{a \dot a}$ from \eqref{3-40}, \eqref{3-44}, and \eqref{3-48} (where $x,y,z,u$ are any combinations of dotted and undotted indices $a, \dot a$).  For $\beta_{a \dot a}, \tilde{\beta}_{a \dot a}, \tilde{\tilde \beta}_{a \dot a}$,
we do the same thing:
\begin{eqnarray}
  \tilde{\tilde \alpha}_{a \dot a} &=&
  \tilde{\alpha}_{a \dot a} - \alpha_{a \dot a},
\label{3-54}\\
  \tilde{\tilde \beta}_{a \dot a} &=&
  \tilde{\beta}_{a \dot a} - \beta_{a \dot a}.
\label{3-58}
\end{eqnarray}
Now we would like to calculate the graded Jacobiator (ternutator):
\begin{equation}
\begin{split}
  J = & \left[
    Q_a, Q_{\dot a}, \left( Q_b Q_{\dot b} \right)
  \right] -
  \left[
    Q_{\dot a}, \left( Q_b Q_{\dot b} \right), Q_{a}
  \right] +
  \left[
    \left( Q_b Q_{\dot b} \right), Q_{a}, Q_{\dot a}
  \right] =
\\
  &
  \left(
    \beta_{b \dot b} - \gamma_{b \dot b} + \tilde \alpha_{b \dot b}
  \right) \left\{ Q_a, Q_{\dot a} \right\} +
  \left(
    \alpha_{a \dot a} - \delta_{\dot a a} + \tilde \beta_{a \dot a}
  \right) \left\{ Q_b, Q_{\dot b} \right\}.
\end{split}
\label{3-51}
\end{equation}
One can see that the Jacobiator will be equal to zero if we choose the following parameters:
\begin{eqnarray}
  \beta_{b \dot b} - \gamma_{b \dot b} + \tilde \alpha_{b \dot b} &=& 0,
\label{3-55}\\
  \alpha_{a \dot a} - \delta_{\dot a a} + \tilde \beta_{a \dot a} &=& 0
\label{3-56}
\end{eqnarray}
but for another choice of these parameters the Jacobian will be nonvanishing.

Now we have to check the consistency of the commutators \eqref{2-30} and \eqref{2-40}. The consistency of the commutator \eqref{2-30} follows from the associator \eqref{3-20}. To check the consistency of the commutator \eqref{2-40}, we substitute $P_\mu$ from \eqref{2-80} into \eqref{2-40}:
\begin{equation}
\label{3-50}
\begin{split}
  \left[ P_\mu, P_\nu \right] = & \frac{1}{16} \sigma_\mu^{a \dot a}
  \sigma_\nu^{b \dot b} \left[
    \left\{ Q_a, Q_{\dot a} \right\}, \left\{ Q_b, Q_{\dot b} \right\}
  \right] =
\\
  &
  \frac{1}{16} \sigma_\mu^{a \dot a}\left(
  \left[
    \left( Q_a Q_{\dot a} \right), \left( Q_b Q_{\dot b} \right)
  \right] +
  \left[
    \left( Q_{\dot a} Q_a \right), \left( Q_b Q_{\dot b} \right)
  \right] +
  \right.
\\
  &
  \left.
  \left[
    \left( Q_{a} Q_{\dot a} \right), \left( Q_{\dot b} Q_b \right)
  \right] +
  \left[
    \left( Q_{\dot a} Q_a \right), \left( Q_{\dot b} Q_b \right)
  \right]
  \right).
\end{split}
\end{equation}
In order to evaluate the right-hand side of \eqref{3-50}, we have to calculate the commutator
\begin{equation}\label{3-60}
\begin{split}
  \left[
    \left( Q_a Q_{\dot a} \right), \left( Q_b Q_{\dot b} \right)
  \right] = & \left( - \alpha_{b \dot b} + \beta_{b \dot b} \right)
  \left\{ Q_a, Q_{\dot a} \right\} +
\\
  &
  \left( \alpha_{a \dot a} - \beta_{a \dot a} \right)
  \left\{ Q_b, Q_{\dot b} \right\} -
\\
  &
  2 \sigma^\alpha_{a \dot b} Q_b \left( P_\alpha Q_{\dot a} \right) +
  2 \sigma^\alpha_{b \dot a} Q_a \left( P_\alpha Q_{\dot b} \right).
\end{split}
\end{equation}
In order to evaluate the right-hand side of \eqref{3-60}, we have employed the following chain:
\begin{equation}
\label{3-62}
\begin{split}
  \left( Q_a Q_{\dot a} \right) \left( Q_b Q_{\dot b} \right)
  & \rightarrow
  Q_a \left( Q_{\dot a} \left( Q_b Q_{\dot b} \right) \right)
  \rightarrow
  Q_a \left( \left( Q_{\dot a} Q_b \right) Q_{\dot b} \right)
  \rightarrow
\\
  &
  Q_a \left( \left( Q_b Q_{\dot a} \right) Q_{\dot b} \right)
  \rightarrow
  Q_a \left( Q_b \left( Q_{\dot a} Q_{\dot b} \right) \right)
  \rightarrow
\\
  &
  \left( Q_a Q_b \right) \left( Q_{\dot a} Q_{\dot b} \right)
  \rightarrow
  \left( Q_b Q_a \right) \left( Q_{\dot a} Q_{\dot b} \right)
  \rightarrow
\\
  &
  Q_b \left( Q_a \left( Q_{\dot a} Q_{\dot b} \right) \right)
  \rightarrow
  Q_b \left( Q_a \left( Q_{\dot b} Q_{\dot a} \right) \right)
  \rightarrow
\\
  &
  Q_b \left( \left( Q_a Q_{\dot b} \right) Q_{\dot a} \right)
  \rightarrow
  Q_b \left( \left( Q_{\dot b} Q_a \right) Q_{\dot a} \right)
  \rightarrow
\\
  &
  Q_b \left( Q_{\dot b} \left( Q_a Q_{\dot a} \right) \right)
  \rightarrow
  \left( Q_b Q_{\dot b} \right) \left( Q_a Q_{\dot a} \right).
\end{split}
\end{equation}
Other commutators can be found by using \eqref{3-60}:
\begin{align}
  \left[
    \left( Q_{\dot a} Q_a \right), \left( Q_b Q_{\dot b} \right)
  \right] =
  \left( \alpha_{b \dot b} - \beta_{b \dot b} \right)
  \left\{ Q_a, Q_{\dot a} \right\} + &
  \left( - \alpha_{a \dot a} + \beta_{a \dot a} \right)
  \left\{ Q_b, Q_{\dot b} \right\} +
\nonumber \\
  2 \sigma^\alpha_{a \dot b} Q_b \left( P_\alpha Q_{\dot a} \right) -
  2 \sigma^\alpha_{b \dot a} Q_a \left( P_\alpha Q_{\dot b} \right) + &
  2 \sigma^\alpha_{a \dot a} \left[
    P_\alpha, Q_b Q_{\dot b}
  \right]
\label{3-70}\\
  \left[
    \left( Q_{a} Q_{\dot a} \right), \left( Q_{\dot b} Q_b \right)
  \right] = \left( \alpha_{b \dot b} - \beta_{b \dot b} \right)
  \left\{ Q_a, Q_{\dot a} \right\} + &
  \left( - \alpha_{a \dot a} + \beta_{a \dot a} \right)
  \left\{ Q_b, Q_{\dot b} \right\} +
\nonumber \\
  2 \sigma^\alpha_{a \dot b} Q_b \left( P_\alpha Q_{\dot a} \right) -
  2 \sigma^\alpha_{b \dot a} Q_a \left( P_\alpha Q_{\dot b} \right) + &
  2 \sigma^\alpha_{b \dot b} \left[
    Q_a Q_{\dot a}, P_\alpha
  \right]
\label{3-80}\\
  \left[
    \left( Q_{\dot a} Q_a \right), \left( Q_{\dot b} Q_b \right)
  \right] = \left( - \alpha_{b \dot b} + \beta_{b \dot b} \right)
  \left\{ Q_a, Q_{\dot a} \right\} + &
  \left( \alpha_{a \dot a} - \beta_{a \dot a} \right)
  \left\{ Q_b, Q_{\dot b} \right\} -
\nonumber \\
  2 \sigma^\alpha_{a \dot b} Q_b \left( P_\alpha Q_{\dot a} \right) +
  2 \sigma^\alpha_{b \dot a} Q_a \left( P_\alpha Q_{\dot b} \right) - &
  2 \sigma^\alpha_{a \dot a} \left[
    P_\alpha \left( Q_b Q_{\dot b} \right) +
    \left( Q_{\dot b} Q_b \right) P_\alpha
  \right] -
\nonumber \\
  2 \sigma^\alpha_{b \dot b} \left[
    \left( Q_a Q_{\dot a} \right) P_\alpha +
    P_\alpha \left( Q_{\dot a} Q_a \right)
  \right] + &
  4 \sigma^\alpha_{a \dot a} \sigma^\beta_{b \dot b}
  \left\{ P_\alpha , P_\beta \right\}.
\label{3-90}
\end{align}
The right-hand side of \eqref{3-70}-\eqref{3-90} are calculated by analogy to \eqref{3-62}.
Substituting \eqref{3-60}-\eqref{3-90} into the right-hand side of~\eqref{3-50}, we obtain the identity
\begin{equation}\label{3-100}
  \left[ P_\mu, P_\nu \right] \equiv \left[ P_\mu, P_\nu \right],
\end{equation}
that proves the consistency of the commutator \eqref{2-40}.

The numbers $\alpha_{a \dot a}$ and $\beta_{a \dot a}$ are still uncertain. Probably the most interesting case is the following:
\begin{eqnarray}
  \alpha_{a \dot a} &=& \zeta \sigma^\mu_{a \dot a} x_\mu,
\label{3-110}\\
  \beta_{a \dot a} &=& - \zeta \sigma^\mu_{a \dot a} x_\mu
\label{3-120}
\end{eqnarray}
where $\zeta$ is some constant. In this case the associator
$\left[ Q_a, Q_{\dot a}, \left( Q_b Q_{\dot b} \right) \right]$ becomes
\begin{equation}\label{3-130}
  \left[ Q_a, Q_{\dot a}, \left( Q_b Q_{\dot b} \right) \right] = 2 \zeta
  \sigma^\mu_{a \dot a} \sigma^\nu_{b \dot b} M_{\mu \nu},
\end{equation}
where the coefficient $\zeta$ equalize the dimensions of the right and left hand sides of \eqref{3-130}. The operator
\begin{equation}\label{3-140}
  M_{\mu \nu} = x_\mu P_\nu - x_\nu P_\mu
\end{equation}
is the angular momentum operator. The inverse relation for \eqref{3-130} is
\begin{equation}\label{3-150}
  M_{\mu \nu} = \frac{1}{8 \zeta}
  \sigma_\mu^{a \dot a} \sigma_\nu^{b \dot b}
  \left[ Q_a, Q_{\dot a}, \left( Q_b Q_{\dot b} \right) \right] .
\end{equation}
Let us check the Nambu fundamental identity
\begin{equation}
\begin{split}
  &\left[ Q_x, Q_y, \left[
      Q_a, Q_{\dot a}, \left( Q_b Q_{\dot b} \right)
    \right]
  \right] = \left[ \left[
      Q_x, Q_y, Q_a
    \right ], Q_{\dot a}, \left( Q_b Q_{\dot b} \right)
  \right] +
\\
  &
  \left[
    Q_a, \left[
      Q_x, Q_y, Q_{\dot a}
    \right], \left( Q_b Q_{\dot b} \right)
  \right] + \left[
    Q_a, Q_{\dot a}, \left[
      Q_x, Q_y, \left( Q_b Q_{\dot b} \right)
    \right]
  \right].
\end{split}
\label{3-200}
\end{equation}
In the consequence of \eqref{2-10}, \eqref{3-20}, \eqref{3-40} and \eqref{3-45} the Nambu fundamental identity \eqref{3-200} has the following form
\begin{equation}
\begin{split}
  &\alpha_{a \dot a} \sigma_{b \dot b}^\mu \left[
    Q_x, Q_y, P_\mu
  \right] +
  \beta_{b \dot b} \sigma_{a \dot a}^\mu \left[
    Q_x, Q_y, P_\mu
  \right] =
\\
  &\begin{cases}
    0,  \quad \text{if $x,y$ are both either dotted or undotted indices}; \\
    \alpha_{c \dot c} \sigma_{b \dot b}^\mu \left[
    Q_a, Q_{\dot a}, P_\mu
  \right] +
  \beta_{b \dot b} \sigma_{c \dot c}^\mu \left[
    Q_a, Q_{\dot a}, P_\mu
  \right],  \quad \text{if } x=c, y=\dot c; \\
  \gamma_{c \dot c} \sigma_{b \dot b}^\mu \left[
    Q_a, Q_{\dot a}, P_\mu
  \right] +
  \delta_{b \dot b} \sigma_{c \dot c}^\mu \left[
    Q_a, Q_{\dot a}, P_\mu
  \right],  \quad \text{if } x = \dot c, y = c .
  \end{cases}
\end{split}
\label{3-210}
\end{equation}
Thus the Nambu fundamental identity is trivial one $0=0$ since $\left[ Q_x, Q_y, P_\mu \right] = 0$, $P_\mu$ is the associative operator.

Let us note that one can generalize the expression \eqref{3-130} in the following way\footnote{thanks for the referee for this comment}
\begin{equation}\label{3-160}
\begin{split}
  \left[ Q_a, Q_{\dot a}, \left( Q_b Q_{\dot b} \right) \right] =
  & a_1 \epsilon_{ab} \epsilon_{\dot a \dot b} \mathbb I +
  \sigma^\mu_{a \dot a} \sigma^\nu_{b \dot b} \left(
    a_2 P_\mu P_\nu + a_3 M_{\mu \nu}
  \right) +
\\
  &
  a_4 \eta_{\mu \nu} \sigma^\mu_{a \dot a} \sigma^\nu_{b \dot b} \mathbb I +
  a_5 \sigma^{\mu \nu}_{a b} \sigma^{\rho \tau}_{\dot a \dot b}
  M_{\mu \nu} M_{\rho \tau} + \cdots 
\end{split}
\end{equation}
where $a_{1,2,\cdots , 5}$ are numerical constants;
$
\sigma^{\mu \nu}_{a b} = \left[
  \sigma^\mu , \sigma^\nu
\right]_{ab}
$,
$
\sigma^{\mu \nu}_{\dot a \dot b} = \left[
  \sigma^\mu , \sigma^\nu
\right]_{\dot a \dot b}
$, and $\epsilon_{ab}, \epsilon_{\dot a \dot b}$ are the antisymmetric symbols.

\section{Decomposition of quantum mechanics operators}

As we saw above, the momentum and the angular momentum operators in quantum mechanics can be decomposed as a multilineal combination of constituents
$Q_a, Q_{\dot a}$,
\begin{eqnarray}
 P_\mu &=& \frac{1}{4} \sigma_\mu^{a \dot a} \left\{
    Q_a, Q_{\dot a}
  \right\},
\label{4-10}\\
  M_{\mu \nu} &=& \frac{1}{8 \zeta}
  \sigma_\mu^{a \dot a} \sigma_\nu^{b \dot b}
  \left[ Q_a, Q_{\dot a}, \left( Q_b Q_{\dot b} \right) \right] .
\label{4-20}
\end{eqnarray}
 It was shown in Ref. \cite{Dzhunushaliev:3221} that the nonrelativistic spin operator $s_i$ can be decomposed as the commutator of octonions (for the definition of octonions and other details, see Appendix \ref{na}),
\begin{equation}\label{4-30}
  s_i = - \frac{1}{4} \epsilon_{ijk} \left[ q_{j+3}, q_{k+3} \right],
  \quad i,j,k =1,2,3,
\end{equation}
where $q_i$ are split-octonions.

All this shows that at least some operators in quantum mechanics can be decomposed as a multilineal combination of constituents.

The decompositions \eqref{4-10}-\eqref{4-30} lead to another interesting question: whether there is a nonassociative algebra $\mathcal A$ in which there exists an associative algebra
$\mathcal Q \subset \mathcal A$ such that $\mathcal Q$ is an algebra of quantum operators?

\section{Discussion and conclusions}

In this Letter we have shown that the idea of supersymmetry can be extended with the inclusion of nonassociativity into supersymmetry. We have defined associators with three and four multipliers and have shown the consistency of these definitions. It is shown that for some special choice of the parameters of four associator such an associator gives rise to an angular momentum operator.

We have seen that momentum, angular momentum, and spin operators have nonassociative decompositions. This allows us to ask the question whether there is a nonassociative algebra $\mathcal A$ in which there exists an associative algebra $\mathcal Q \subset \mathcal A$ such that $\mathcal G$ is the algebra of quantum operators. This means that the operators $G \in \mathcal G$ are the operators of either quantum mechanics or quantum field theory. If the  answer is positive, then an interesting situation occurs: any quantum operator can be decomposed as a multilineal combination of nonassociative constituents. In such a situation one can give a positive answer to the old question whether there is an alternative quantum mechanics. The fact of the matter is that one can rearrange brackets in the nonassociative operator decomposition of quantum operator and redefine the action for the wave-function operator.

In connection with the decompositions \eqref{4-10}-\eqref{4-30}, one can remember the hidden variables theory (HVT) where hidden variables are the classical ones. The HVT argues that a quantum state of a physical system does not give a complete description of the system.  It is assumed in the HVT that quantum mechanics represents a statistical approximation of
an unknown deterministic theory, where all observables have defined values
fixed by unknown variables. The difference between the decompositions \eqref{4-10}-\eqref{4-30} presented here and the HVT is that constituents $Q_a, Q_{\dot a}$ are nonassociative quantities but in the HVT the unknown variables are classical ones. Moreover, one can show \cite{Dzhunushaliev:2007vg} that nonassociative quantities are unobservable ones. This means that nonassociative decompositions
cannot be considered as the HVT.

\subsection*{Acknowledgment}
This work is supported by a grant of the VolkswagenStiftung and by a grant in fundamental research in natural sciences by the Ministry of Education and Science of Kazakhstan. Also I am grateful to V. Folomeev for the fruitful discussion and the referee for useful comments.

\appendix
\section{Octonions}
\label{na}

The split-octonions are nonassociative numbers. The split-octonions have the following commutators and associators:
\begin{eqnarray}
 \left[ q_{i+3} , q_{j+3} \right] &=& - 2 \epsilon_{ijk} q_k ,
\label{app-10}\\
 \left[ q_i, q_j \right] &=& 2 \epsilon_{ijk} q_k  ,
\label{app-20}\\
 \left( q_{i+3}, q_{j+3}, q_{k+3} \right) &=& \left( q_{i+3} q_{j+3} \right) q_{k+3} -
 q_{i+3} \left( q_{j+3} q_{k+3} \right) = 2 \epsilon_{ijk} q_7,
\label{app-30}
\end{eqnarray}
where $i,j,k = 1,2,3$. The commutator \eqref{app-20} shows that $q_i$ (with $i=1,2,3$) form a subalgebra. This subalgebra is called the quaternion algebra $\mathbb H$; $q_{1,2,3}$ are quaternions. The commutator \eqref{app-20} is the same as the commutator relationship for spin operators $s_i = \sigma_i$  ($\sigma_i$ are the Pauli matrices). The relations \eqref{app-10}-\eqref{app-30} can be written in the Zorn vector matrix representation, where the octonion $o$ is written in the form
\begin{equation}\label{app-40}
  o = \sum \limits_{i=1}^8 \alpha_i q_i =
  \left(
	\begin{array}{cc}
		a				&	\vec x	\\
		\vec y	&	b
	\end{array}
	\right),
\end{equation}
where $q_0 = 1$; $\alpha_i$ are numbers; $a,b$ are real numbers; and $\vec x, \vec y$ are 3-vectors. The product of two octonions is defined as
\begin{equation}	
	\left(
	\begin{array}{cc}
		a				&	\vec x	\\
		\vec y	&	b
	\end{array}
	\right)
	\left(
	\begin{array}{cc}
		c				&	\vec u	\\
		\vec v	&	d
	\end{array}
	\right) =
	\left(
	\begin{array}{rl}
		ac + \vec x \cdot \vec v										&	\quad a \vec u + d \vec x -
		\vec y \times \vec v 																														 \\
		c \vec y + b \vec v + \vec x \times \vec u	&	\quad bd + \vec y \cdot \vec u
	\end{array}
	\right)
\label{app-50}
\end{equation}
here $( \cdot )$ and $[ \times ]$ denote the usual scalar and vector products. Let us introduce the orthogonal vectors $\vec e_i$, where $i=1,2,3$, such that $\vec e_i \times \vec e_j = \epsilon_{ijk} \vec e_k$ and
$\vec e_i \cdot \vec e_j = \delta_{ij}$. Then the split-octonions have the following Zorn vector matrix representation:
\begin{equation}	
	1 = \left(
	\begin{array}{ll}
		1				&	\vec 0	\\
		\vec 0	&	1
	\end{array}
	\right), 
	q_7 = - \left(
	\begin{array}{cc}
		1				&	\vec 0	\\
		\vec 0	&	-1
	\end{array}
	\right), 
	q_i = \left(
	\begin{array}{ll}
		0					&	- \vec e_i	\\
		\vec e_i	&	0
	\end{array}
	\right), 
	q_{i+3} = \left(
	\begin{array}{cc}
		0					&	\vec e_i	\\
		\vec e_i	&	0
	\end{array}
	\right),
\label{app-60}
\end{equation}
where $i=1,2,3.$ Thus, the nonrelativistic spin operators have two matrix representations: the first one is the Pauli matrix representation with
$s_i = \frac{\sigma_i}{2}$, and the second one is the Zorn vector matrix  representation \eqref{app-60} with $s_i = \frac{\imath}{2} q_i, i=1,2,3$.


\begin{thebibliography}{1}

\bibitem{okubo1995}
Susumu Okubo, \textit{Introduction to Octonion and Other Non-Associative Algebras in Physics}, Cambrodge University Press, Cambridge, 1995.

\bibitem{Carrion:2003ve}
  H.~L.~Carrion, M.~Rojas and F.~Toppan,
  JHEP {\bf 0304}, 040 (2003);
  [hep-th/0302113].

\bibitem{Castro:2012zzb}
  C.~Castro,
  Int.\ J.\ Geom.\ Meth.\ Mod.\ Phys.\  {\bf 9}, 1250021 (2012).

\bibitem{Gogberashvili:2005xb}
  M.~Gogberashvili,
  J.\ Phys.\ A {\bf 39}, 7099 (2006);
  [hep-th/0512258].

\bibitem{Gogberashvili:2005cp}
  M.~Gogberashvili,
  Int.\ J.\ Mod.\ Phys.\ A {\bf 21}, 3513 (2006);
  [hep-th/0505101].

\bibitem{Beggs:2005zt}
  E.~J.~Beggs and S.~Majid,
  J.\ Math.\ Phys.\  {\bf 51}, 053522 (2010);
  [math/0506450 [math-qa]].

\bibitem{Sabinin:2001pd}
  L.~V.~Sabinin,
  Int.\ J.\ Theor.\ Phys.\  {\bf 40}, 351 (2001).

\bibitem{Wulkenhaar:1996at}
  R.~Wulkenhaar,
  Phys.\ Lett.\ B {\bf 390}, 119 (1997);
  [hep-th/9607096].

\bibitem{fsusy}
Ahn C.  Bernard D. and  Leclair A.,
{\it  Nucl. Phys. B },  \textbf{34}   409-439 (1990).

\bibitem{Mohammedi:2003qx}
  N.~Mohammedi, G.~Moultaka and M.~Rausch de Traubenberg,
  Int.\ J.\ Mod.\ Phys.\ A {\bf 19}, 5585 (2004)
  [hep-th/0305172].

\bibitem{Rausch de Traubenberg:2003bi}
  M.~Rausch de Traubenberg,
  eConf C {\bf 0306234}, 578 (2003)
  [hep-th/0312066].

\bibitem{Bars:1978yx}
  I.~Bars and M.~Gunaydin,
  J.\ Math.\ Phys.\  {\bf 20}, 1977 (1979).

\bibitem{Castro:2014tda}
  C.~Castro,
  Int.\ J.\ Geom.\ Meth.\ Mod.\ Phys.\  {\bf 11}, no. 3, 1450013 (2014).

\bibitem{Yamazaki:2008gg}
  M.~Yamazaki,
  Phys.\ Lett.\ B {\bf 670}, 215 (2008)
  [arXiv:0809.1650 [hep-th]].

\bibitem{Dzhunushaliev:3221}
V.~Dzhunushaliev,
J. Gen. Lie Th. Appl., \textbf{3}, 33-38 (2009);
[arXiv:0805.3221].

\bibitem{Dzhunushaliev:2007vg}
  V.~Dzhunushaliev,
  J.\ Math.\ Phys.\  {\bf 49} (2008) 042108
  [arXiv:0712.1647 [quant-ph]].
\end{thebibliography}
\end{document}